\documentclass[preprint2]{aastex}

\shorttitle{TYPE DEPENDENCE IN THE $\mu_0$--$\log h$ PLANE}
\shortauthors{GRAHAM \& DE BLOK}

\begin{document}

\input{psfig}

\title{A MORPHOLOGICAL-TYPE DEPENDENCE IN THE $\mu_{0}$--$\log h$ PLANE OF SPIRAL GALAXY DISKS}
\author{Alister W. Graham}
\affil{Instituto de Astrof\'{i}sica de Canarias, La Laguna,
E-38200, Tenerife, Spain}
\email{agraham@ll.iac.es}

\and 

\author{W.J.G. de Blok\altaffilmark{1}}
\affil{ATNF, P.O.\ Box 76, Epping, NSW 1710, Australia}
\email{edeblok@atnf.csiro.au} 

\altaffiltext{1}{Bolton Fellow}

\begin{abstract}

We present observational evidence for a galaxy `Type' dependence to the 
location of a spiral galaxy's disk parameters in the $\mu_{0}$--$\log h$ 
plane.  With a sample of $\sim$40 Low Surface Brightness galaxies 
(both bulge- and disk-dominated) and $\sim$80 High Surface Brightness 
galaxies, the early-type disk 
galaxies ($\leq$Sc) tend to define a bright envelope in the 
$\mu_0$--$\log h$ plane, while the late-type ($\geq$Scd) spiral galaxies 
have, in general, smaller and fainter disks.  Below the defining surface 
brightness threshold for a Low Surface Brightness galaxy (i.e.\ more 
than 1 mag fainter than the 21.65 B-mag arcsec$^{-2}$ Freeman value), 
the early-type spiral galaxies have scale-lengths greater than 8-9 kpc, 
while the late-type spiral galaxies have smaller scale-lengths. 
All galaxies have been modelled with a seeing-convolved S\'ersic $r^{1/n}$ 
bulge and exponential disk model.  We show that the trend of decreasing 
bulge shape parameter ($n$) with increasing Hubble type and decreasing 
bulge-to-disk luminosity ratio, which has been observed amongst the 
High Surface Brightness galaxies, extends to the Low Surface Brightness 
galaxies, revealing a continuous range of structural 
parameters.  

\end{abstract}


\section{Introduction}

During the closing decade of the 1900s considerable progress was 
made in the observation and parameterization of intrinsically faint 
spiral galaxies (van der Hulst et al.\ (1993); 
de Blok et al 1995; Sprayberry et al.\ 1995; Bothun, 
Impey, \& McGaugh 1997; Beijersbergen, de Blok, \& van der Hulst 1999). 
Galaxies with a central disk surface brightness more than one magnitude 
fainter than the canonical Freeman (1970) value of 21.65 B-mag arcsec$^{-2}$
were defined to be Low Surface Brightness (LSB) galaxies.   Spiral 
galaxies with a brighter central disk surface brightness have 
subsequently been labelled High Surface Brightness (HSB) galaxies, 
and it is these galaxies which typify the Hubble sequence. Although, 
in reality, this threshold is somewhat arbitrary as observations 
of many spiral galaxies reveals a smooth continuation in central disk 
surface brightness values across this boundary.  
The existence of LSB galaxies -- including knowledge of the 
extreme LSB galaxy `Malin 1' (Bothun et al.\ 1987, Impey \& Bothun 
1989) -- had of course been known many years earlier (Longmore et al.\ 
1982; Davies et al.\ 1988), and had been previously predicted, 
on the grounds of visibility arguments Disney (1976).  

In this Letter we combine the data from two LSB galaxy samples
with a representative sample of HSB galaxies containing all spiral 
morphological types.  The objective is to explore the tentative suggestion 
of Graham (2001a) that a morphological type dependence exists in the 
central disk surface brightness -- disk scale-length ($\mu_0$--$\log h$) 
diagram.  
Re-analyzing the light-profiles from the diameter-limited sample of 86
spiral galaxies from de Jong \& van der Kruit (1994) and de Jong (1996), 
Graham (2001a) noted that the low surface brightness (high $\mu_0$) 
small disk scale-length corner of the $\mu_0$--$\log h$ 
diagram was dominated, in fact, only populated, by late-type spiral
galaxy disks.  To be somewhat more quantitative, the host galaxies of 
disks with scale-lengths less than 9 kpc and central surface brightnesses 
fainter than 19.0 K-mag arcsec$^{-2}$ were all late-type spiral galaxies 
($>$Sc).  To investigate this further, this same galaxy sample is
supplemented here with two of the largest samples of LSB galaxies 
measured with CCDs.  The first is a sample of $\sim$20 `disk-dominated' LSB 
galaxies from de Blok et al.\ (1995), and the second is a sample of 
$\sim$20 `bulge-dominated' LSB spiral galaxies (Beijersbergen et al.\ 1999).

The selection criteria for the three data samples are reviewed 
in Section 2, and the derivation of the light-profile model 
parameters is also discussed there. 
All of the galaxies have been modelled by us using the same bulge/disk 
decomposition algorithm.  We additionally provide the first quantitative
measurement to the prominence, and/or absence, of the bulges in 
`disk-dominated' LSB galaxies. 
The resulting distribution of disks in the
$\mu_0$--$\log h$ diagram is presented in Section 3.  

\section{Data}

\subsection{The galaxy sample}

The data selection procedure, and method of reduction, are described 
in detail in the above mentioned papers.  We will only comment here 
on the selection criteria -- relevant to concerns about possible 
biases which might result in an under-estimation of early-type 
spirals in the low surface brightness (high $\mu_0$) small 
scale-length corner of the $\mu_0$--$\log h$ diagram.  

de Jong \& van der Kruit (1994) selected `undisturbed' spiral galaxies 
from the UGC catalog (Nilson 1973) if they had a red minor-axis to
major-axis ratio greater than 0.625 (i.e.\ inclinations less than 
$\sim$50$^{\rm o}$), a red major-axis $\geq$2.0\arcmin, and an 
absolute galactic latitude greater than 25$^{\rm o}$.  Telescope 
time and pointing restrictions resulted in a final sample of 83 
spiral galaxies with types ranging from Sa to Sm (plus one S0 and 
two Irregular galaxies).  None of their selection criteria are 
expected to create a morphological-type bias.  

Beijersbergen et al.\ (1999) selected spiral 
galaxies from the ESO-LV catalog (Lauberts \& Valentijn 1989) 
to have types ranging from Sa-Sm, inclinations less than 50$^{\rm o}$, 
and absolute galactic latitudes greater than 15$^{\rm o}$.  They 
additionally required 
the diameters of the 26 B-mag arcsec$^{-2}$ isophote to be greater than 
1\arcmin and smaller than 3\arcmin (chip size limitation), and the 
surface brightness of the disk at the 
half-light radii to be fainter than 23.8 B-mag arcsec$^{-2}$.  
From the resulting list of some 600 galaxies, a
random sample which had central light concentrations was selected. 
This prescription certainly should not have biased against the 
selection of early-type spiral galaxies, but on the contrary selected 
them if they do indeed exist.  

de Blok et al.\ (1995) selected galaxies from the late-type sample 
of LSB UGC galaxies from van der Hulst et al.\ (1993) and the lists by 
Schombert \& Bothun (1988) and Schombert et al (1992).  Of the 17
galaxies from the de Blok sample for which we have B-band data, 2
originate from the van der Hulst UGC sample.  The others are from 
the Schombert LSB galaxy lists which were constructed from a visual 
inspection of sky survey 
plates, and contain no {\it a priori} bias against early-type spiral 
galaxies. de Blok et al.\ randomly selected a sub-set of galaxies with 
inclinations less than 60$^{\rm o}$, central surface 
brightnesses fainter than 23 B-mag arcsec$^{-2}$, and having 
single-dish H$_{\rm I}$ observations available. 
This last criteria is the only possible bias we can identify (given 
that some early-type spiral galaxies may be gas-poor) ; however,
it is expected to be a small effect for the following reason.
The morphological make-up of the LSB galaxy catalogs of 
Schombert et al.\ (consisting of objects not in the UGC) turned out to 
be dominated (85\%) by late-type galaxies (another 10\% 
are dwarf ellipticals and some 5\% unclassifiable).  The reason for 
this composition is that the prominence of the bulge in early-type 
spiral galaxies had resulted in their prior detection in the 
first-generation Palomar sky-survey plates used by Nilson.  The 
second-generation sky-survey plates went about a magnitude deeper, 
and predominantly resulted in the detection of fainter late-type 
galaxies.  Thus, even if the H$_{\rm I}$ requirement excluded {\it all} 
early-type galaxies in the LSB catalogs, this would only be a 5-10\%
bias, which would equate to the exclusion of 1 or 2 early-type galaxies 
from our sample.  

To summarize, the total galaxy sample is likely to be biased in favor 
of the inclusion of early-type LSB galaxies due to the selection 
criteria used by Beijersbergen et al.\ (1999).

\subsection{The light-profile model parameters}

The galaxy light-profiles from the three galaxy samples are 
already published in the respective papers, and we refer the interested
reader to these.  For the sake of the present analysis, 
we have applied the same algorithm to every light-profile, 
simultaneously fitting a seeing-convolved S\'ersic (1968) $r^{1/n}$ 
bulge and a seeing-convolved exponential disk (see Graham 2001b 
for details).   This is an improvement 
over previous comparative work between the HSB and LSB galaxies 
because it not only avoids biases between different authors 
fitting techniques, but provides a more reliable approach to 
separating the bulge light from that of the disk.  Additionally, this 
is the first time a bulge model has been fitted to the `disk-dominated' 
sample of LSB galaxies.
The structural parameters obtained from this decomposition are: the
central disk surface brightness ($\mu_0$), the disk scale-length ($h$),
the effective radius of the bulge ($r_e$), and the surface brightness
of the bulge at this radius ($\mu_e$), and lastly, the shape parameter
$n$ from the $r^{1/n}$ bulge model. 

After correcting the heliocentric velocity measurements 
for Virgo-infall using the 220 model of Kraan-Korteweg (1986), 
the scale-lengths were converted from arcseconds to kpc 
using a Hubble constant $H_0$$=$75 km s$^{-1}$ Mpc$^{-1}$. 
Redshifts for ESO-LV 1590200, ESO-LV 0350110, and ESO-LV 0050050 
were obtained from the Parkes telescope HIPASS database 
(Barnes et al.\ 2000).  Redshifts for the other galaxies came from 
the respective papers.  
The surface brightnesses were adjusted for 
a) $(1+z)^4$ redshift dimming, 
b) $K$-corrections (using the tables of Poggianti 1997), 
c) Galactic extinction (Schlegel, Finkbeiner, \& Davis 1998; 
data obtained from NED), and 
d) inclination corrected.  We used the inclination correction 
2.5$C$$\log(a/b)$, where $a/b$ is the axis-ratio of the outer
disk -- assumed to be due to inclination.  The value of $C$ 
was taken to be 0.2 for the $B$-band and 0.5 for the $R$-band (Tully
\& Verheijen 1997).  It is noted that while this correction 
is necessary for the HSB galaxies, it's relevance to the LSB galaxies 
is not as clearly established.  Both the HSB and the LSB galaxy 
samples have inclinations less than 60$^{\rm o}$, and so the maximum 
correction is 0.15 mag arcsec$^{-2}$ ($B$-band) and $\sim$0.4 
mag arcsec$^{-2}$ ($R$-band).  The average correction for the 
LSB galaxies turned out to be only 0.06 ($B$-band) and 0.15 ($R$-band) mag 
arcsec$^{-2}$, and cannot be a significant bias or explanation for the 
distribution of points in the $\mu_0$--$\log h$ diagram (Section 3). 

We report that only a couple of the late-type LSB galaxies from de Blok 
et al.\ (1995) 
have bulges with central surface brightness values that are brighter 
than the disk, and even then, only brighter by a few tenths of 
magnitude.  Most of the late-type LSB galaxies have bulges with central 
surface brightness values that are fainter than that of the disk; their 
presence signalled by a central bulge in the light-profile that 
rises less than $0.75(=2.5\log(2))$ mag (and often much less) above the 
extrapolated disk light-profile.   Many of the light-profiles actually 
show no sign at all of a bulge.  
In all the late-type LSB galaxies in the sample, the light distribution 
is overwhelmingly dominated by that of the disk; however, for those LSB 
galaxies with bulges, we can perform a comparison study with the HSB 
galaxy sample.

\section{The $\mu_0$--$\log h$ diagram}

\begin{figure}
\plottwo{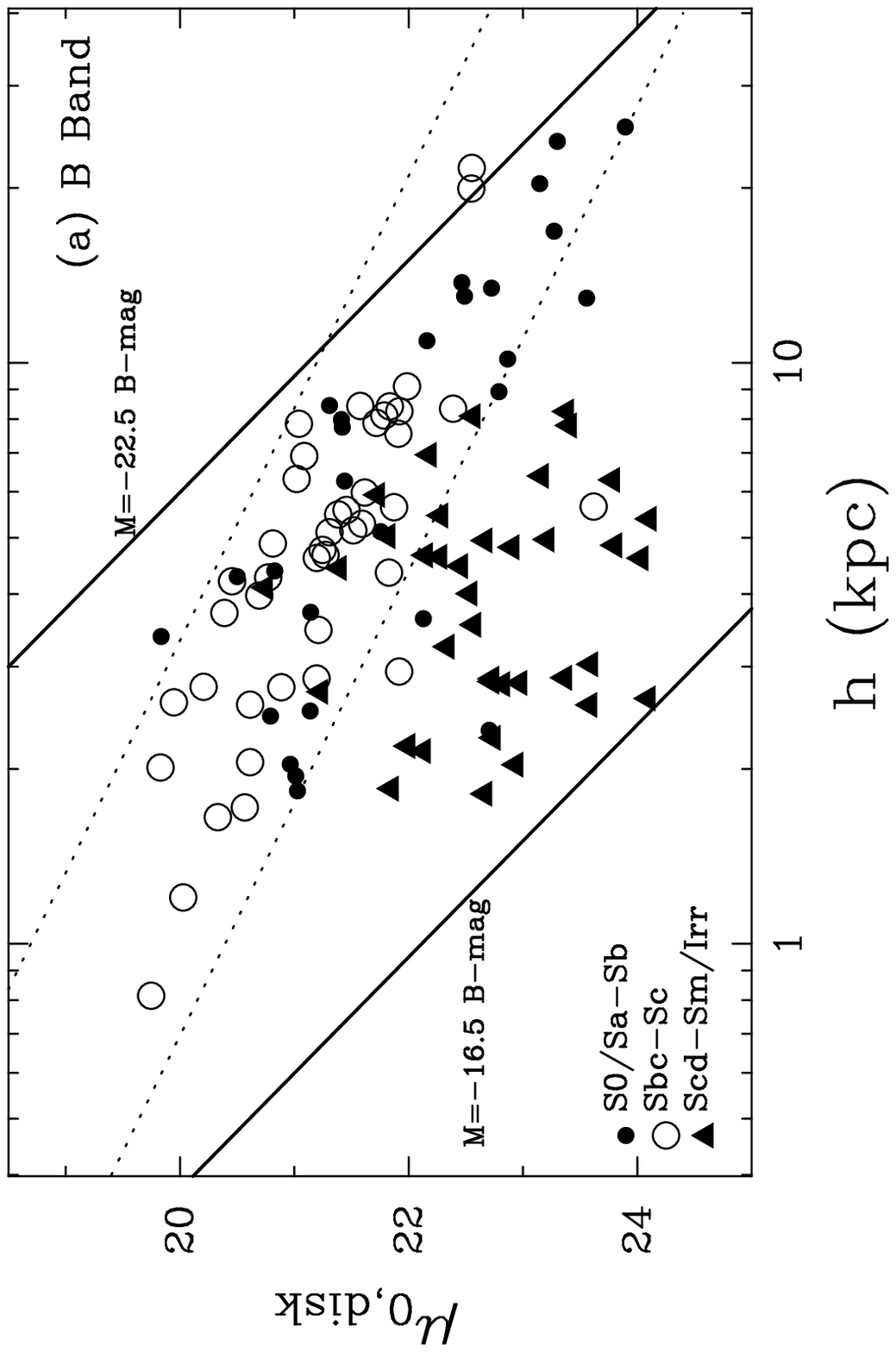}{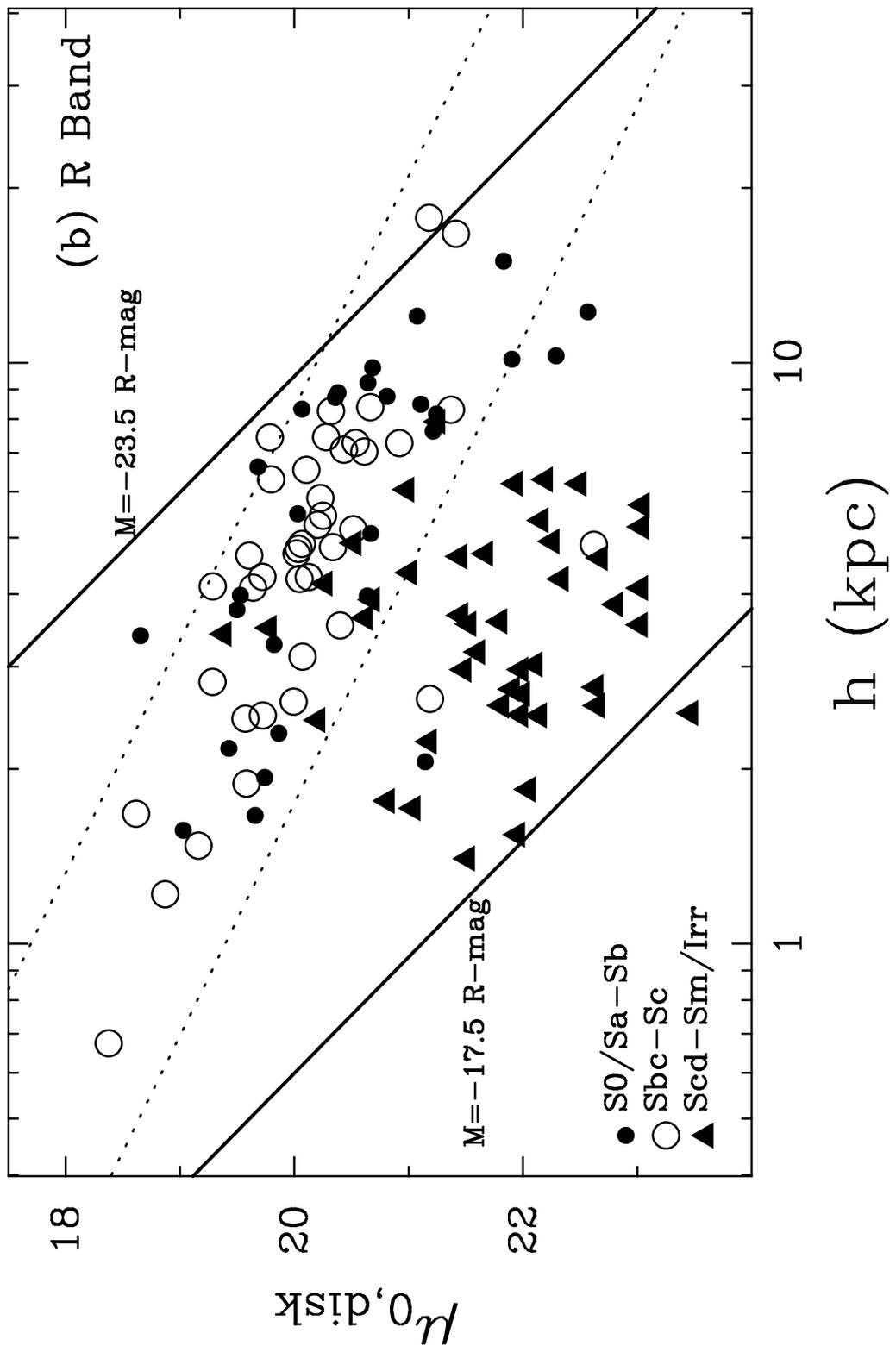}
\caption{
(a) The $B$-band central disk surface brightnesses are plotted against
the disk scale-lengths ($H_0$=75 km s$^{-1}$ Mpc$^{-1}$).
The solid lines are lines of constant disk luminosity (slope equals 5),
and the dotted lines have a slope of 2.5.  
Galaxies of different morphological type are marked with 
different symbols, according to the legend inset in the diagram. 
(b) The $R$-band data.
}
\label{fig1}
\end{figure}

Shown in Figure~\ref{fig1} are the central disk surface brightnesses
plotted against the associated disk scale-lengths. In general, the disks 
from early-type spiral galaxies tend to reside, or rather, define, the
upper envelope of points in the $\mu_0$--$\log h$ diagram.  The disks of 
late-type spiral galaxies tend to reside, in general, below this 
envelope.  
One can also get a feel for the distribution of points by cutting the 
diagram up further.  While we do not wish to suggest anything 
physical with an arbitrary cut at 22.65 B-mag arcsec$^{-2}$, 
it is however somewhat revealing to do so.   
One can see that the early-type spiral galaxies with central disk 
surface brightnesses fainter than this value have large disk scale-lengths
($>$8-9 kpc) while the LSB late-type spiral galaxies have disk scale-lengths 
less than this.  Yet another way to view the diagram is parallel
to lines of constant disk luminosity, which have a slope of 5.  
One can see that, on average, 
the brightest early-type spiral galaxies have brighter luminosities than the
brightest late-type spirals; a result consistent with the luminosity 
functions of different spiral galaxy types (Folkes et al.\ 1999). 
Furthermore, when one factors in the contribution from the bulge
light this trend will be stronger still.  The data does however reveal 
that there is some overlap of galaxy types in the diagram.  
That is, rather than a strict exclusion of a given population in 
different parts of the $\mu_0$--$\log h$ diagram. it is more a 
case of one population dominating over the other.  
This morphological type separation is actually analogous to the 
behavior of bulges in the $\mu_e$--$\log r_e$ diagram (Kent 1985; 
his figure 5b). 

Past studies of the $\mu_0$--$\log h$ diagram have either been 
limited to the upper right of the diagram by the selection-criteria, 
or have not investigated the morphological-type dependence.  
Bothun et al.\ 1997 explored beyond the typical selection-criteria 
boundaries by including in their Figure 4, in addition to a sample 
of HSB galaxies, a large sample of LSB galaxies.  
They revealed a similar distribution of 
points as seen in Figure~\ref{fig1}, but no morphological-type 
distinction was made (similarly see, Dalcanton, Spergel, \& 
Summers 1998 figure 4; McGaugh 1998 figure 1). 
In reviewing the literature, however, we have found that combining 
figure 3 from Tully \& Verheijen (1997) (a $\mu_0$--$\log h$ diagram) 
with their figure 6 (a plot of $\mu_0$ vs.\ galaxy type) 
should reveal the same morphological-type dependence as observed here, 
lending support to this result. 

\begin{figure}
\plottwo{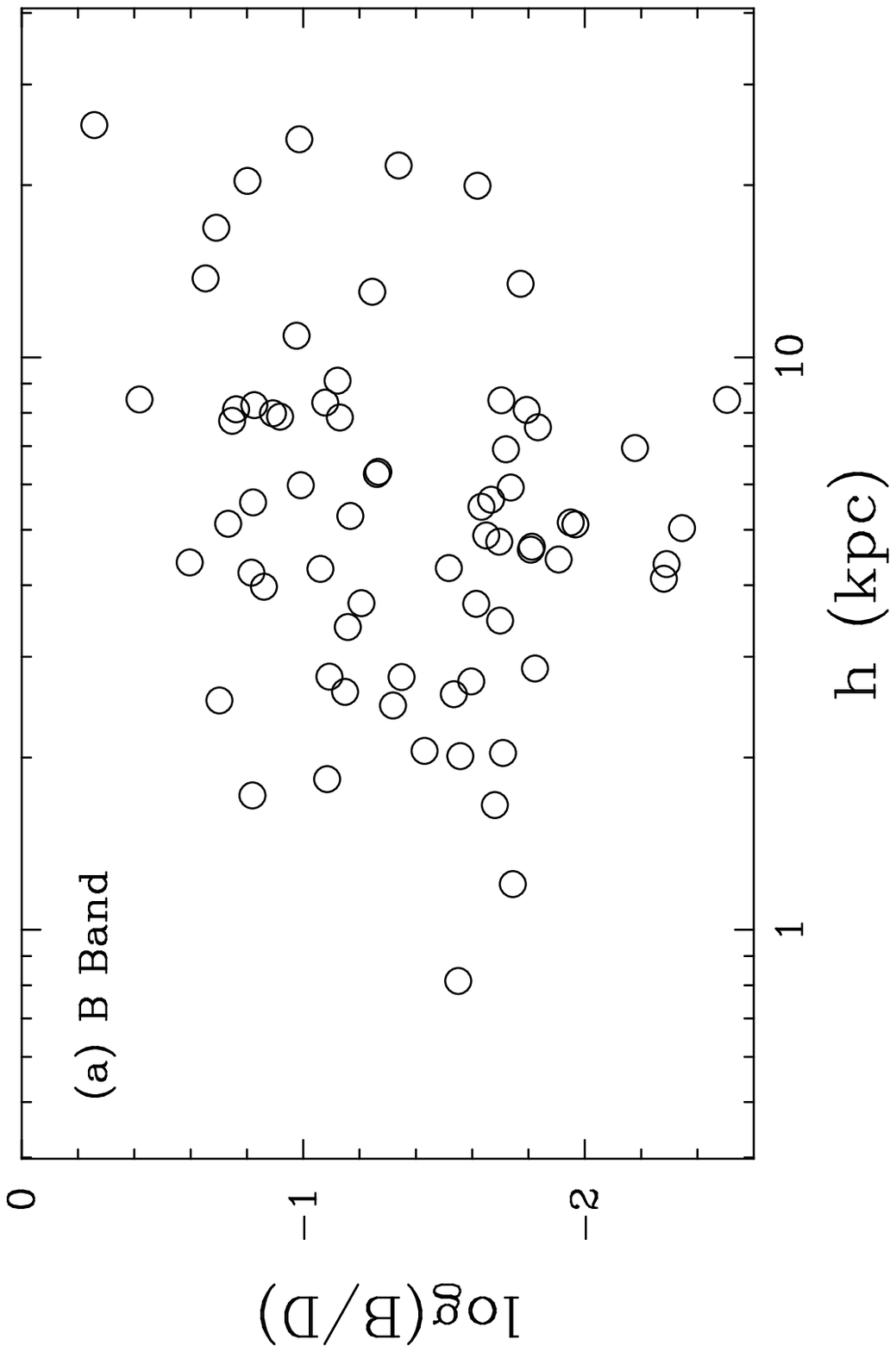}{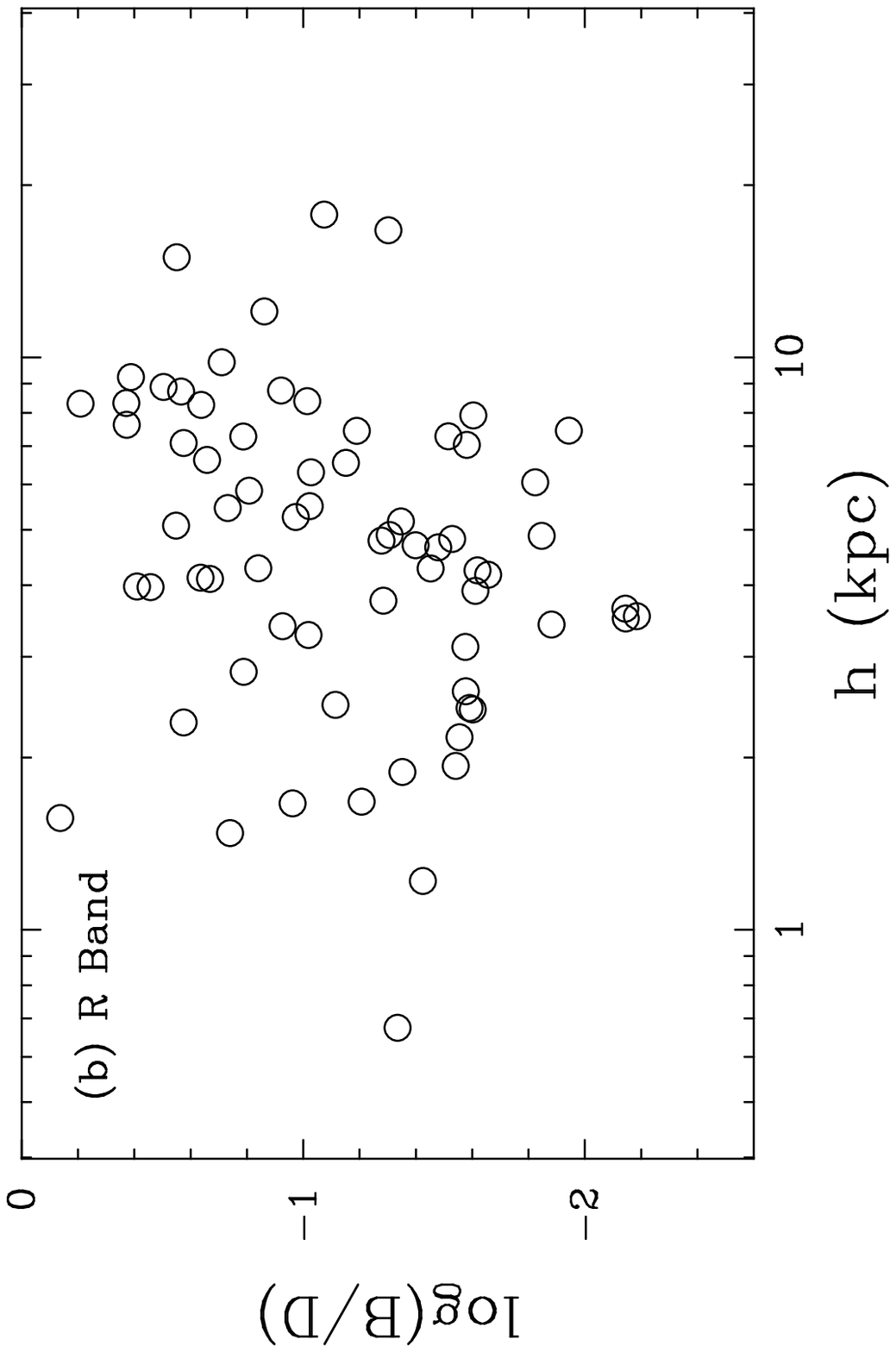}
\caption{
(a) The $B$-band bulge-to-disk luminosity ratios from galaxies in the
upper envelope of Figure~\ref{fig1} (see text for details) are plotted
against their associated disk scale-length.
(b) The $R$-band data.
}
\label{fig2}
\end{figure}

The upper boundary in the $\mu_0$--$\log h$ plane 
represents a real decline to the space-density of galaxies with big 
(large $h$) bright (low $\mu_0$) disks -- the selection criteria 
favors their detection.  The presence of this boundary has been used
in Graham (2001a, see also Graham 2000a,b) to estimate the degree of
opacity in inclined early-type spiral galaxy disks; inclined galaxies 
with transparent disks are shown to reside above this envelope due 
to the apparent brightening of their disk with inclination. 
The lower boundary is however artificial -- the result of a cut-off 
imposed by the galaxy selection criteria . 
The existence of ultra-low LSB, optically invisible galaxies 
(Disney 1998), is therefore not excluded.  
The (solid) lines of constant luminosity shown in Figure~\ref{fig1} are 
lines of constant disk luminosity, and therefore do not include 
possible contributions from the bulge light to the total galaxy light.
One may therefore wonder if the upper envelope may in fact represent 
a boundary of constant `galaxy' luminosity -- perhaps the bright 
end to the luminosity function.  To address this thought, we show 
in Figure~\ref{fig2} the bulge-to-disk ($B/D$) luminosity ratio, 
as a function of disk 
scale-length for every galaxy in the upper envelope delineated by 
the dotted lines in Figure~\ref{fig1}.  No trend of increasing $B/D$ 
luminosity ratio with decreasing scale-length is evident, and 
consequently no support is given to the suggestion that the upper 
boundary is due to a maximum galaxy luminosity. 

\begin{figure}
\plottwo{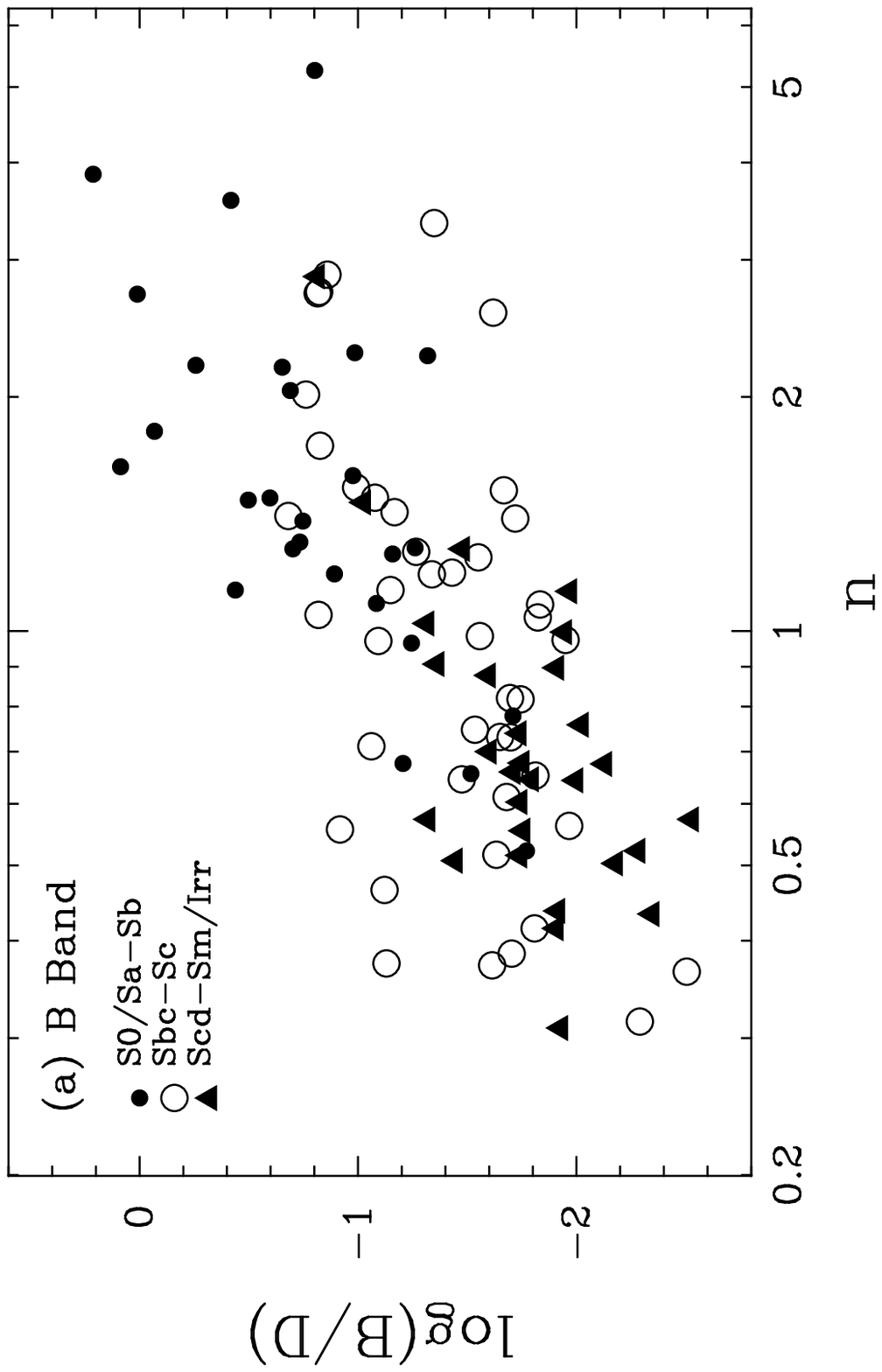}{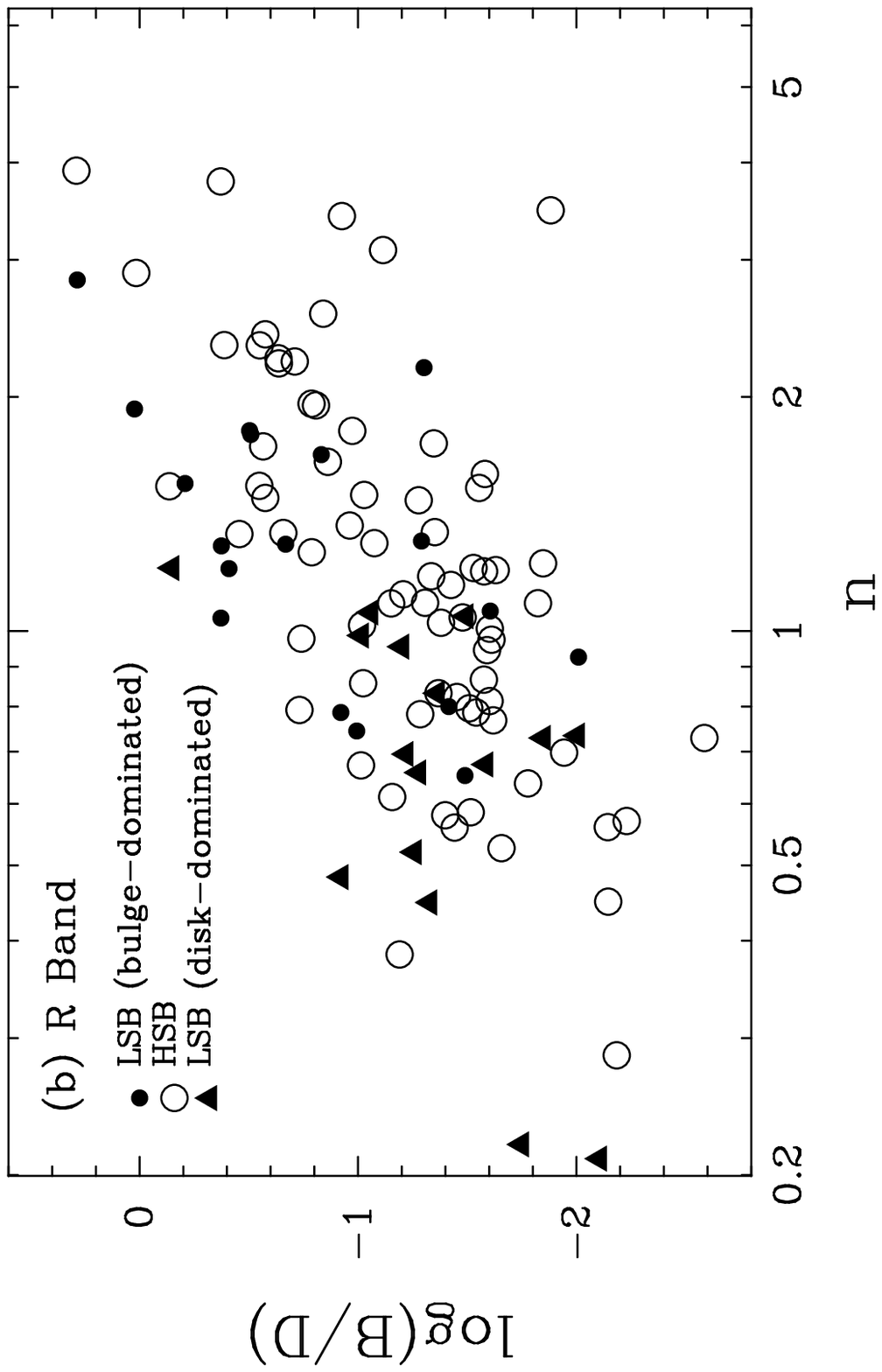}
\caption{
(a) The $B$-band bulge-to-disk luminosity ratios are plotted as a function of
the $B$-band $r^{1/n}$ bulge shape parameter $n$.  The points are labelled
according to the host galaxy morphological type.  
(b) The $R$-band bulge-to-disk luminosity ratio are plotted as a function of
the $R$-band $r^{1/n}$ bulge shape parameter $n$.  The points are labelled
according to the galaxy sample they came from.  
(Galaxies with no discernible bulge are not plotted.) 
}
\label{fig3}
\end{figure}

In Figure~\ref{fig3} we have plotted all the $B/D$ luminosity ratios 
against the bulge shape parameter $n$ (the index from the best-fitting 
$r^{1/n}$ bulge profile models).  These points have been plotted 
as a function of galaxy type, and also in a way so as to reveal 
which galaxy sample they have come from.  One can see that for both
the HSB and LSB galaxies, 
the shape of the bulge light-profile is correlated with the nature of
the host galaxy in the sense that both the $B/D$ luminosity 
ratio and the galaxy morphological type are correlated with $n$, 
This result is firmly established for the bulges of HSB spiral 
galaxies (Andredakis, Peletier, \& Balcells 1995; Graham 2001b; 
Mollenhoff \& Heidt 2001), but has not been previously shown 
for the bulges of LSB spiral galaxies.  It would appear that no 
obvious structural distinction can be made between the LSB and 
HSB spiral galaxies in this diagram.

\acknowledgments
We wish to thank Ren\'{e}e Kraan-Korteweg for making her   
Virgo-centric inflow code available. 
This research has made use of the NASA/IPAC Extragalactic Database (NED)
which is operated by the Jet Propulsion Laboratory, California Institute
of Technology, under contract with the National Aeronautics and Space 
Administration. 
The Parkes telescope is part of the Australia Telescope which is funded 
by the Commonwealth of Australia for operation as a National Facility 
managed by CSIRO.

\clearpage

\end{document}